\documentstyle[12pt,epsf]{article}
\begin{document}
\textwidth 13.6cm
\textheight 19.9cm
\noindent {\large \bf Polymer adsorption on curved surface} \\

\noindent {\it \large Pramod Kumar Mishra} \\
\noindent {\bf Department of Physics, DSB Campus, Kumaun University \\
Naini Tal-263 002, India} \\

\noindent {\bf Abstract} :

Lattice model of directed self avoiding walk is used to investigate adsorption 
properties of a semiflexible sequential copolymer chain on an impenetrable curved surface
on a hexagonal lattice in two dimensions. 
Walks of the copolymer chains are directed in a direction away from the surface
at a suitable value of monomer surface attraction, 
the copolymer chain gets adsorbed on the surface. To calculate exact value
of monomer surface attraction, the directed walk model have been solved
analytically using generating function method to discuss results when one type
monomer of the copolymer chain has attractive, repulsive or no interaction with
the surface. Results obtained show that 
adsorption transition point is independent of bending energy
of the copolymer chain. 

\vspace {.2cm}
\noindent {\bf PACS Nos.: 05.70.Fh, 64.60 Ak, 05.50. +q}
\section{Introduction:}
The copolymer adsorption is a subject of theoretical and experimental investigations
due to its applications in biophysics, biosensors, pattern recognition, adhesion
and surface protection. For example, in the case of biophysics, Since $protein$ molecule is made
of heterogeneous sequence of hydrophobic and hydrophilic residues, therefore, it can be treated 
as a random copolymer and its stiffness may be in between flexible
and stiff chains. Whence, $protein$ is a semiflexible random copolymer molecule.     
The conformational properties of such polymer molecules have attracted considerable attention
in recent years due to advancement in the experimental methods in which it has become 
possible to pull and stretch single bio-molecule to measure its elastic properties 
$\cite{1,2}$. These study reveal a wealth of information about the conformational 
behaviour of biopolymers and therefore of biological importance. 
In addition to it, study of adsorption of the copolymer chain 
on a surface is also useful in determining 
the relation between their compositions with their adsorption characteristics.

The problem of random copolymer adsorption has been extensively studied
using numerical methods, see, $\cite{3,4,5,6,7,8}$ and references quoted therein. 
However, analytical methods for
adsorption of copolymer chain with self avoidance effect are limited to 
directed walk like models. 
In past few years sequential copolymer adsorption has also received attention $\cite{9}$
due to location of its adsorption transition point and calculation 
of crossover exponent. 
Adsorption of
the copolymer chain merits somewhat differently from homopolymer chain 
in a sense that different type of monomers of the copolymer chain need not have attractive
interaction with the surface.    

Lattice model of self avoiding walks and directed self avoiding walks 
have been used extensively to derive essential physics associated with the behaviour of a surface
interacting polymer chain in a good solvent in two and three dimensions $\cite{9,10,11,12}$.
If surface is attractive, it contributes an energy $\epsilon_s$
($<0$) for each step of walk lying on the surface. This leads
to an increased probability defined by a Boltzmann weight
$\omega=\exp(-\epsilon_s/k_BT)$ of moving a step on the surface
($\epsilon_s < 0$ or $\omega > 1$, $T$ is temperature and
$k_B$ is the Boltzmann constant). 
The polymer
chain gets adsorbed on the surface at a suitable value of $\epsilon_s$. 
The transition
between adsorbed and desorbed regimes is marked by a critical value of adsorption
energy or $\omega_c$. 
One may define the crossover
exponent $\phi$ at the adsorption transition point as, $N_s \sim N^{\phi}$, where $N$ is the total
number of monomers in the polymer chain and $N_s$ the number of monomers adsorbed on the surface. 

In this paper, we have extended directed self avoiding walk model, introduced
by Privman {\it et al.} $\cite{12}$ for homopolymer chain, to study the 
adsorption desorption phase transition behaviour of the semiflexible sequential copolymer chain 
immersed in a good solvent on a curved (one dimensional) impenetrable surface.
To calculate adsorption properties of the copolymer chain on a curved impenetrable surface,
we have considered two dimensional hexagonal lattice. 
Such study is useful in examining the question whether the 
copolymers differ from homopolymers with respect to their critical behaviour near an impenetrable surface.
We have considered semiflexible sequential copolymer chain composed of two type of monomers (A \& B)
and model the copolymer chain as a directed self avoiding walk on the lattice.
Such copolymer model serve as a paradigmatic model of actually disordered macromolecules
(for example, $proteins$).
For adsorption of semiflexible sequential copolymer chain on an impenetrable flat and curved surface 
perpendicular to the preferred direction of the walks of the copolymer chain,   
directed walk model has been solved analytically and exact critical value of the surface attraction
for the adsorption of the copolymer chain has been evaluated. 

The paper is organized as follows: In Sec. 2 lattice model of
directed self avoiding walk has been described for the semiflexible sequential copolymer chain 
in two dimensions on a hexagonal
lattice. Directed self avoiding walk models of the copolymer chain has been solved analytically for the adsorption
of the chain on an impenetrable curved surface. 
Finally, in Sec. 3 we discuss the results obtained.

\section{Model and method}
A lattice model of directed self-avoiding walk $\cite{12}$ has been used to 
study adsorption-desorption phase transition behaviour of a sequential copolymer chain under 
good solvent condition on a curved impenetrable surface. 
The directed walk models are
restrictive in the sense that angle of bending has a unique value that is 
$60^{\circ}$ for each bend. 
In addition to it, directedness of the walk amounts to some degree of stiffness in the 
copolymer chain because all
directions of the space are not treated equally. 
However, directed self avoiding walk model can be solved 
analytically and therefore gives exact value of  
adsorption transition point of the chain.  

For two dimensional hexagonal lattice, we have consider directedness
of the copolymer chain that can be defined with the help
of direction of movement of the walker on the unit cell of
the lattice as follows:
There are six possible directions of movement
of the walker and steps along these directions can be named by 1, 2, 3, 4, 5 and 6 (as shown in Fig. 1).
Time direction is along step 6.
If walker is allowed to take steps along all the possible
directions excluding only directions along step 2, 3 \&4. 

\begin{figure}[htbp] 
\centering 
\epsfxsize=5cm\epsfbox{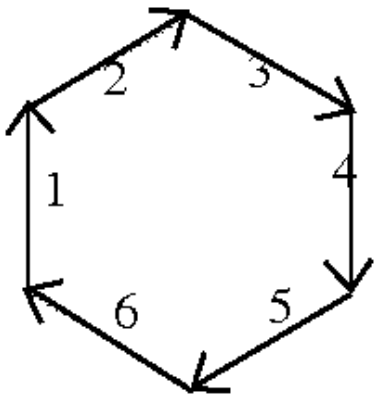} 
\caption{ All the possible directions of movement of the walker on a two dimensional hexagonal
lattice has been named by 1, 2, 3, 4, 5 and 6.}
\end{figure}

The stiffness in the sequential copolymer chain has
been introduced by associating an energy barrier for each bend in the walk of copolymer chain. 
The stiffness weight $k=exp(-\beta\epsilon_{b})$ where $\beta=(k_BT)^{-1}$ is 
inverse of the temperature and $\epsilon_b(>0)$ is the energy associated with 
each bend of the walk of copolymer chain. For $k=1$ or $\epsilon_{b}=0$ the copolymer chain is said to be flexible and for 
$0<k<1$ or $0<\epsilon_{b} <\infty$ the copolymer chain is said to be 
semiflexible. However, when $\epsilon_{b}\to\infty$ or $k\to0$, 
the copolymer chain has a rigid rod like shape.

The partition function of a semiflexible sequential copolymer chain made of two type of monomers (A \& B) can be written as, 

\begin{equation}
Z(k,x_1,x_2)={\sum}^{N=\infty}_{N=0}\sum_{ all\hspace{0.07cm}walks\hspace{0.07cm}of\hspace{0.05cm}N\hspace{0.05cm}steps} {x_1}^{N/2}{x_2}^{N/2}k^{N_b}
\end{equation}
where, $N_b$ is the total number of bends in a walk of $N$ steps (monomers),
$x_1$ and $x_2$ is the step fugacity of each of the A and B type monomers respectively.
For the sake of mathematical simplicity we assume here onwards $x_1=x_2=x$. Method
of analysis discussed in this paper can be easily extended to the case $x_1\ne x_2$. 

\subsection{Polymer chain in the bulk: Directed self avoiding walk model on a two dimensional hexagonal lattice}
$Y(k,x)$ is sum of Boltzmann weight of all the walks having first step along step 1 or step 5 while $X(k,x)$ is sum of Boltzmann weight of all the
walks having first step along step 6. One end of the chain grafted at $O$, as shown in figure (2).
The components of partition function for copolymer chain in the bulk can be written as (as shown in figure 2B),

\begin{equation}
Y(k,x)=x+kx^2+2k^2x^2Y
\end{equation}
and,
\begin{equation}
X(k,x)=x+2xkY
\end{equation}
solving Eqs. (2 \& 3), we have,
\begin{equation}
Y(k,x)=-\frac{x+x^2k}{-1+2k^2x^2}
\end{equation}
\begin{equation}
X(k,x)=-\frac{x+2kx^2}{-1+2k^2x^2}
\end{equation}
With $O$ as starting point where chain is grafted, so that partition function of the chain $G(k,x)$ can be written as,
\begin{equation}
G(k,X)=2Y(k,x)=-\frac{2(x+x^2k)}{-1+2k^2x^2}
\end{equation}
from singularity of the partition function $G(k,x)$, we have $x_c$=$\frac{1}{\sqrt{2}k}$, required for polymer polymerization of 
an infinitely long linear polymer chain.
\begin{figure}[htbp] 
\centering 
\epsfxsize=14cm\epsfbox{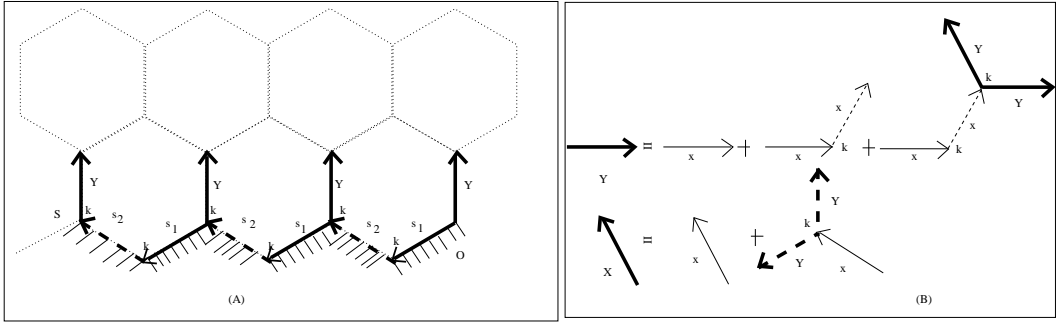} 
\caption{In this figure surface interacting copolymer chain is shown for directed self avoiding walk model on the hexagonal lattice.
Walks of the chain starts from at $O$ from curved surface. $S(k,\omega_1,\omega_2,x)$ is the surface component of the partition function
of the polymer chain.}
\end{figure} 
\subsection{Adsorption on a curved surface: Directed self avoiding walk model on a two dimensional hexagonal lattice}
Adsorption of the copolymer chain has been studied on a two dimensional hexagonal lattice to investigate 
adsorption desorption phase transition behaviour of a semiflexible sequential copolymer chain on a curved surface. 
In the case of a two dimensional hexagonal lattice,
surface is an impenetrable line and its shape is like a saw tooth wave $\cite{14}$. 
Therefore, adsorbed parts of the copolymer chain have bends and in this case components of
partition function of the copolymer chain having first monomer
grafted to the surface of A type can be written following the method outlined above as,
 
\begin{equation}
S(k,\omega_1,\omega_2,x)=s_1(1+ks_2+k^2s_2Y)+{s_1}^2s_2k^2(1+ks_2+k^2s_2Y)+\dots
\end{equation}
$S(k,\omega_1,\omega_2,x)$ and $Y(k,x)$ are the components of the partition function
on the surface and perpendicular to the surface respectively.
\begin{equation}
S(k,\omega_1,\omega_2,x)=\frac{s_1(1+ks_2+k^2s_2Y)}{1-s_1s_2k^2}\hspace{2cm} (s_1s_2k^2<1)
\end{equation}

The partition function of surface interacting copolymer chain can be written for 
the sequential copolymer chain for two
dimensional hexagonal lattice as follows:

The component of the partition function perpendicular to
the plane of the surface is $\cite{14}$,
\begin{equation}
Y(k,x)=-\frac{x+kx^2}{-1+2x^2k^2}
\end{equation}
so that, 
\begin{equation}
\hspace{-2cm}G_s(k,\omega_1,\omega_2,x)=S(k,\omega_1,\omega_2,x)+Y(k,x)
\end{equation}
\begin{equation}
\hspace{-2cm}G_s(k,\omega_1,\omega_2,x)=\frac{f(k,\omega_1,\omega_2,x)}{(-1+k^2s_1s_2)(-1+2k^2x^2)} \hspace{.5cm}(s_1s_2k^2<1)
\end{equation}

In this case, singularities of the partition function give, 
$\omega_{c1}=\frac{2}{\omega_{c2}}$.

Variation of $\omega_{c1}$ for various value of $\omega_{c2}$ is shown in Fig. (3) 
for two dimensional hexagonal lattice. When we
substitute $\omega_{c1}=\omega_{c2}=\omega_c$, we are able to obtain critical value
of monomer surface attraction required for adsorption of homopolymer chain
on a curved impenetrable surface. It is to be
noted that $\omega_{c1}$ required for adsorption of copolymer chain on a curved
impenetrable surface is independent of bending energy of the copolymer chain. 
\begin{figure}[htbp] 
\centering 
\epsfxsize=9cm\epsfbox{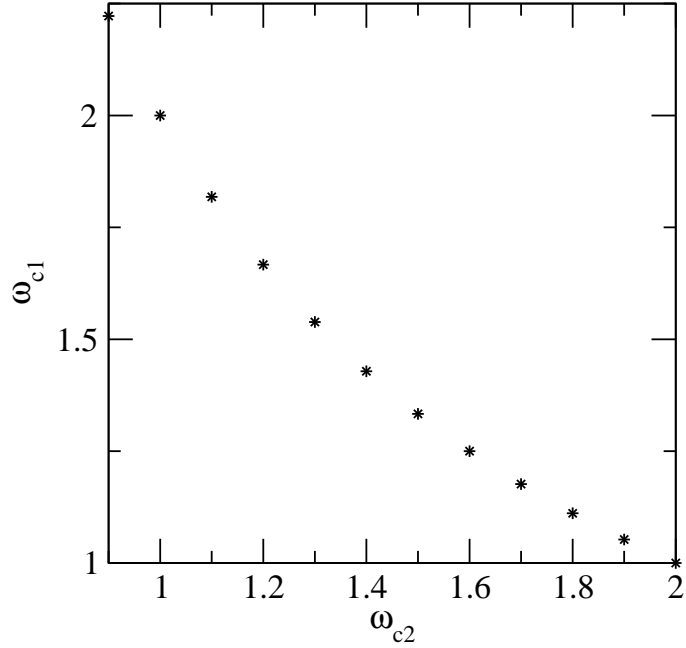} 
\caption{Wariation of $\omega_{c1}$ with $\omega_{c2}$ is shown for hexagonal lattice. }
\end{figure} 
\section{Result and discussion}
Lattice model of directed self avoiding walk has been solved
using generating function method for adsorption of a semiflexible sequential copolymer chain 
on a curved surface. We have used two dimensional hexagonal lattice to model the copolymer chain
and to investigate adsorption properties of the copolymer chain on an impenetrable
curved surface. 
The copolymer chain is made of two type of monomers (A \& B) and 
A type monomer has attractive interaction with the surface while B type monomer
can have attractive, repulsive or no interaction with the surface.
Our study revealed that for curved impenetrable surface, critical value of
monomer surface attraction required for the copolymer
chain adsorption is independent of bending energy
of the chain \cite{14}.

\end{document}